\documentclass[runningheads]{llncs}
\usepackage{graphicx}
\usepackage{cite}
\usepackage{amsmath}
\usepackage{amsfonts}

\usepackage{xr}

\makeatletter
\newcommand*{\addFileDependency}[1]{%
  \typeout{(#1)}
  \@addtofilelist{#1}
  \IfFileExists{#1}{}{\typeout{No file #1.}}
}
\makeatother

\newcommand*{\myexternaldocument}[1]{%
    \externaldocument{#1}%
    \addFileDependency{#1.tex}%
    \addFileDependency{#1.aux}%
}
\myexternaldocument{Supplementary}

\usepackage{xcolor}

\begin{document}
\title{LiftReg: Limited Angle 2D/3D \\ Deformable Registration}

\author{Lin Tian\inst{1} \and Yueh Z. Lee\inst{2} \and Raúl San José Estépar\inst{3} \and Marc Niethammer\inst{1}}

\authorrunning{L. Tian et al.}

\institute{Department of Computer Science, University of North Carolina at Chapel Hill \and
Department of Radiology, University of North Carolina at Chapel Hill \and
Harvard Medical School}
\maketitle              %
\begin{abstract}
We propose LiftReg, a 2D/3D deformable registration approach. LiftReg is a deep registration framework 
which is trained using sets of digitally reconstructed radiographs (DRR) and computed tomography (CT) image pairs. By using simulated training data, LiftReg can use a high-quality CT-CT image similarity measure, which helps the network to learn a high-quality deformation space. To further improve registration quality and to address the inherent depth ambiguities of very limited angle acquisitions, we propose to use features extracted from the backprojected 2D images and a statistical deformation model. We test our approach on the DirLab lung registration dataset and show that it outperforms an existing learning-based pairwise registration approach.

\end{abstract}
\section{Introduction}
The goal of 2D/3D medical image registration is to estimate a spatial transformation between a patient's 3D image volume and one or multiple 2D images under patient deformation. It is widely used in radiation therapy and for image-guided interventional procedures to obtain patient motion, e.g., to compute dose accumulation for multi-fraction radiotherapy, or device locations~\cite{jaffray2007review}. In such applications, Computed Tomography (CT) is a commonly used 3D imaging modality and radiographs are generally adopted to obtain 2D projection images because of their comparatively low radiation dose and fast acquisition speed~\cite{markelj2012review}. In practice, only a limited number of radiographs can be acquired, typically within a limited angle range, resulting in a lack of spatial information in the projection direction and in consequence in spatial ambiguities. Large non-rigid motions (e.g., between lung inspiration/expiration) make registration in this setting particularly challenging.

The conventional approach for 2D/3D deformable image registration (DIR) is to numerically solve an optimization problem to determine the transformation parameters which best explain the deformation between a CT image and a corresponding set of 2D radiographs~\cite{flach2014deformable,prummer2006multi,zikic2008deformable,tian2020fluid}. Image similarity is assessed between the 2D radiographs and the respective simulated CT projections, i.e., the digitally reconstructed radiographs (DRR)~\cite{sherouse1990computation}. However, given very limited angle radiograph acquisitions, assessing image similarity in such a way is susceptible to spatial ambiguity due to the lack of spatial information in the projection direction.

Compared to the optimization-based 2D/3D DIR methods, current learning-based approaches replace optimization by prediction at test-time and and can therefore significantly reduce registration runtime~\cite{zhang2021unsupervised,pei2017non, foote2019real, li2020non}. Zhang~\cite{zhang2021unsupervised} proposes 2D3D-regnet which uses a U-Net as the backbone network and directly predicts deformation vector fields (DVF) given a CT image and a set of radiographs. Evaluation is on a lung dataset~\cite{zhang2016biomechanical} with small deformations (mean landmark distances of $6.5mm$) and results in a mean target registration error (mTRE) of $4.3mm$ after registration for a scanning angle of $30^\circ$. This work computes the similarity loss in 2D space, thus it faces the aforementioned spatial ambiguity issues. Other work~\cite{pei2017non,li2020non} leverages prior information during training to restrict deformations to a reasonable transformation space. Specifically, this stream of work proposes building a parametric deformation subspace with respect to an atlas image via principal component analysis (PCA) computed from a prior cohort of data from which deformation fields are extracted. Though greatly reducing the complexity of the problem, the spatial ambiguity issue of the 2D similarity measure is still not addressed. Foote et al~\cite{foote2019real} build a subject-specific deformation subspace and train a subject-specific network with the ground truth coefficients computed from 3D image pairs of the same subject. While this approach does not suffer from the spatial ambiguity issue, the approach assumes that training data and test data come from the same patient and multiple 3D images of the same patient are available during training. Thus, the approach is not directly applicable to finding deformations between one CT image and a set of radiographs.

The critical question for 2D/3D DIR is then: what kind of information can we leverage to address the spatial ambiguity? One advantage of deep learning is that extra information can assist the learning without changing the problem setting during inference. Thus, leveraging precise 3D information in the loss function during training is expected to reduce spatial ambiguity while retaining generalizability at test time when 3D image pairs are not available. However, this information has not been used for the current existing methods. In addition, how much spatial information the features extracted from 2D images can convey also plays an important role. Compared to the 2D features adopted in~\cite{pei2017non,li2020non,foote2019real}, our network can extract 3D spatial information from a multi-channel backprojected volume. Moreover, we can further reduce the complexity brought by the spatial ambiguity and large freedom of transformation model via prior knowledge of the motion as used in the previously discussed atlas registration approaches. 

We propose LiftReg for 2D/3D registration. Our contributions include:
\begin{itemize}
\item[1)] We propose Lift3D, a module to backproject 2D images to 3D. We experimentally show that Lift3D helps improve 2D/3D registration performance.

\item[2)] We propose LiftReg, a pairwise 2D/3D learning based DIR framework which leverages the precise spatial information in DRR-CT pairs via a \emph{3D} CT-CT similarity loss and by using a statistical deformation model for image motion.   

\item[3)] We demonstrate that LiftReg outperforms an existing learning based 2D/3D pairwise DIR method~\cite{zhang2021unsupervised} on the public DirLab lung dataset with a limited scanning angle of $30^\circ$ for both coarse and fine structures (e.g. lung vessels).
\end{itemize}

\section{Problem formulation}\label{sec:problem_formulation}
Let the two images $I_s:\Omega_s\to{\mathbb{R}}$ and $I_t:\Omega_t\to{\mathbb{R}}$ defined in 3 dimensional space $\Omega\subset{\mathbb{R}^3}$ represent the source and target images and $\textbf{P}_{t}=\{P_i|i=1...N,P_i:\mathbb{R}^2\to\mathbb{R}\}$ be the set of 2D projections of $I_t$. Our goal is to find the spatial mapping $\varphi:\Omega_s\to\Omega_t$ which makes the warped source image $I_s\circ\varphi^{-1}$ consistent with the projections $\textbf{P}_t$. One non-parametric model for $\varphi^{-1}$ is the deformation vector field (DVF), $u(x):\Omega_t\to\mathbb{R}^3$, which  captures the displacement between corresponding points in the target and source domains: $\varphi^{-1}(x)=x+u(x)$.

\begin{figure}[htb]
\vspace{-2mm}
    \centering
    \includegraphics[width=0.854\linewidth]{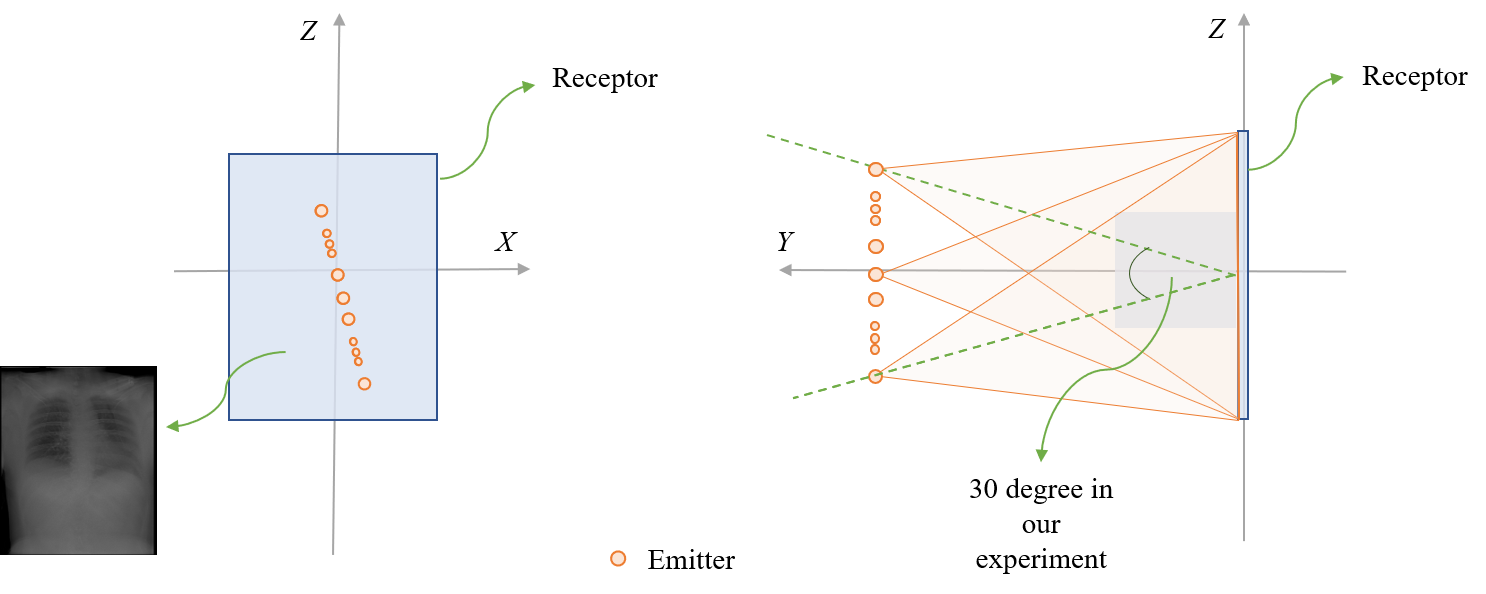}
    \caption{$\textbf{P}_t$ \textbf{acquisition geometry~\cite{Shan2015}.} Left: geometry in top-down view. Emitters are evenly distributed along a line which is not necessarily aligned with the axis of the receptor. Right: side view of the geometry where the patient is in between the emitters and the receptor. The emitters are activated consecutively to acquire $\textbf{P}_t$.}
    \label{fig:DRR_geometry}
    \vspace{-4mm}
\end{figure}
The projections $\textbf{P}_t$ can be acquired based on different geometries: e.g., parallel beam, fan beam and cone beam geometries~\cite{kalender2006x}. In this work, we follow the stationary chest tomosynthesis (s-DCT) imaging geometry in \cite{Shan2015} where several emitters are evenly placed along a line parallel to the receptor covering a limited angle. The receptor stays stationary while the emitters are activated consecutively to obtain the individual projections. Fig.~\ref{fig:DRR_geometry} shows the s-DCT geometry. 

\section{Method}

\subsection{PCA-based deformation vector field subspace}\label{sec:dvf_subspace}

Our first goal is to obtain a PCA-based 3D deformation model. To this end, we compute the gold standard DVFs, $u(x)$, of the training dataset of 3D CT image pairs $(I_s, I_t)$ via an in-house trained 3D/3D deep registration network. Note that other deep registration networks could be used to obtain the gold standard DVFs. The in-house 3D/3D deep registration network is used only to obtain registrations fast. We use singular value decomposition on the displacement fields to obtain the eigenvectors $E=\{e_i|e_i\in\mathbb{R}^s, s=D\times{H}\times{W}\times{3}, i=1...N_e\}$ which explain $99\%$ of the displacement field variance which form our parametric DVF subspace. 

In contrast to atlas and group-wise registration~\cite{pei2017non,foote2019real,li2020non} where the transformation model is defined in a common atlas space, in our case the transformation model DVF, $u(x)$, is defined in the target domain, $\Omega_t$, which is different for each image pair $(I_s,I_t)$. Hence, the deformation space needs to account for deformations within and across patients. To alleviate this issue we only focus on the deformation driven by the region of interest (ROI), namely the lung region. Specifically, we train the 3D/3D deep registration network guided by a similarity loss computed on the ROI only. Thus, it focuses the deformation prediction inside the lung. In addition, the images are processed so that the ROI is centered in the image volume. With the constructed subspace, the transformation model can be expressed as
\begin{equation}\label{equ:deformation_model}
    \varphi^{-1}(x) = x+\bar{u}+\sum_{i=1}^{N_e}\alpha_i{e_i}\,,
\end{equation}
where $\alpha_i$ are the basis coefficients, and $\bar{u}$ indicates the mean DVF.

\begin{figure}[h]
\vspace{-2mm}
    \centering
    \includegraphics[width=1\linewidth,trim={0 2.5cm 0 0},clip=true]{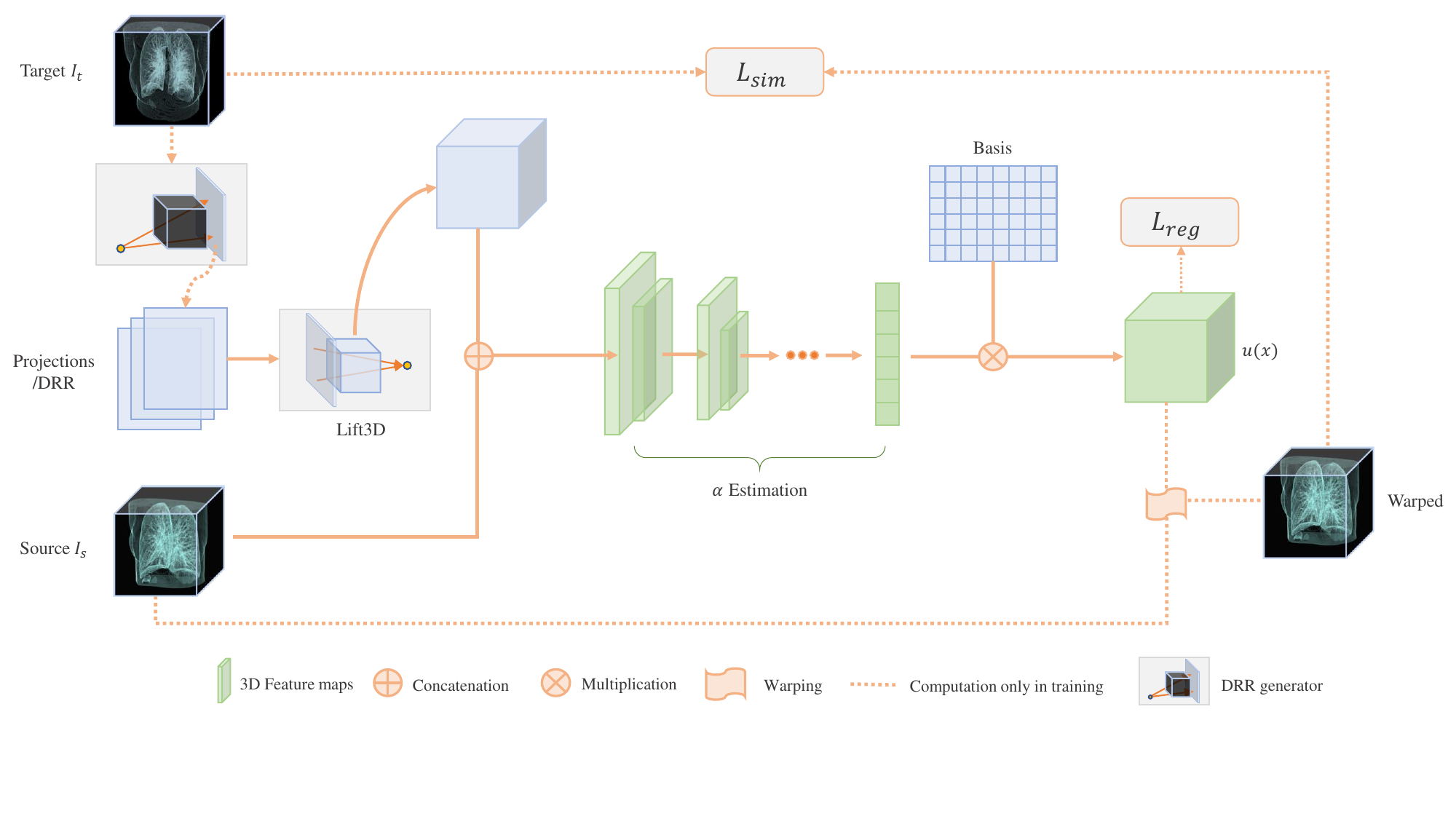}
    \caption{\textbf{Network structure for LiftReg.} The network contains three modules: Lift3D, $\alpha$ estimation, and a warping module. Lift3D backprojects the 2D images to 3D space. $\alpha$ estimation contains several convolutional blocks and fully connect layers to predict the coefficients of the DVF subspace. The warping module warps the source image $I_s$ based on the predicted $\varphi^{-1}(x)$. The computation graph for training is denoted as an orange dotted line. Learnable modules are shown in green. The 3D feature maps are drawn as 2D feature maps for simplicity.}
    \label{fig:network_structure}
    \vspace{-8mm}
\end{figure}

\subsection{Network structure}

We use a convolutional neural network (CNN) $\mathcal{F}_\theta$ to predict the basis coefficients $\{\alpha_i\}$ given a 3D source image and a set of target projections $(I_s, \textbf{P}_t)$: $\{\alpha_i\} = \mathcal{F}_\theta(I_s, \textbf{P}_t)$.
The network $\mathcal{F}_\theta$ contains three major modules: Lift3D, $\alpha$ estimation, and a warping module. The network first backprojects $\textbf{P}_t$ to 3D space based on the associated geometry in which the $\textbf{P}_t$ are acquired. The $\alpha$ estimation module takes as input the lifted 3D image and $I_s$ to predict the $\{\alpha_i\}$. The corresponding transformation $\varphi^{-1}$ is computed via Eq.~\eqref{equ:deformation_model}.
Fig.~\ref{fig:network_structure} shows an overview of $\mathcal{F}_\theta$. Next, we introduce each module in detail.

{\bf Lift3D} Since the radiographs $\textbf{P}_t$ are in 2D space, it is difficult to extract 3D information using a 2D feature extractor. Hence, we lift the 2D images to 3D image space. Specifically, we  backproject the 2D images based on their associated emitter positions. For each pixel in $P_i$, there is a backprojection ray associated to its emitter position. We assign the intensity of the pixel in $P_i$ to all the voxels along the ray. Thus, we obtain a set of 3D volumes $\{I_t^i|i=0...N\}$. We then concatenate this set of 3D images into one $N$ channel 3D volume of size of ${D}\times{W}\times{H}$. This approach has two  benefits: (1) geometry information is now explicitly contained in the input to the network; (2) the visual differences across $N$ lifted 3D volumes can provide visual 3D features for the following network modules. For practical purposes, we implement Lift3D using the following discretization approach. We transform the voxel position $x$ in 3D to its corresponding pixel position in $P_i$ and assign the interpolated intensity in $P_i$ to $x$. Given the emitter positions $\{c_i=(c_i^0, c_i^1, c_i^2)|i=0...N,c_i\in\mathbb{R}^3\}$, we construct the transformation matrix $T$ to transform a 3D coordinate to a 2D homogeneous coordinate in the projection space,
which is similar to the intrinsic matrix used in a pinhole camera model. Given $T$, we can compute the intensity at position $x\in{\Omega_t}$ as 
\begin{equation}
    I_t(x) = P_i(Tx), T=\begin{bmatrix}
    c_i^1 & 0 & c_i^0 \\
    0 & c_i^1 & c_i^2 \\
    0 & 0 & 1
    \end{bmatrix}
\end{equation}
where $c_i^j$ is the $jth$ component of emitter position $c_i$.

{\bf $\alpha$ estimation} This module predicts the coefficients $\{\alpha_i\}$ given $I_s$ and the lifted 3D volumes $\{I_t^i|i=0,...N\}$. We first concatenate $\{I_t^i|i=0,...N\}$ and $I_s$ along the channel dimension. We then pass the concatenated multi-channel 3D volume to multiple layers of convolution blocks where each block contains one convolutional layer, one activation layer and a pooling layer. Lastly, we flatten out the feature maps and apply three fully connect layers to predict the $\{\alpha_i\}$. The detailed structure is described in the supplementary.

{\bf Warping module} After obtaining $\{\alpha_i\}$, we compute the DVF via Eq.~\eqref{equ:deformation_model} and apply the transformation to $I_s$ using a spatial transformer~\cite{jaderberg2015spatial}. 

\subsection{Network training}
Current learning-based 2D/3D DIR methods train their models using a CT/DRR dataset and 2D projection losses. They do not use the 3D volume associated with the DRRs. Instead, we use it to obtain precise 3D information available during training, but not needed at test time, to help network training. 

We use two loss terms: a 3D similarity loss and a regularizer on the deformation field. The overall loss is  
\begin{equation} \label{equ:problem_definition}
       \mathcal{L} = {\mathcal{L}}_{sim}+\lambda{\mathcal{L}}_{reg}\,,
\end{equation}
where $\lambda>0$. We use Normalized Cross-correlation (NCC) as the similarity measure. To help the network focus on information within the lung ROI, we mask the lung region based on the pre-computed segmentation maps $S_s$ and $S_t$ (for $I_s$ and $I_t$ respectively) before computing the similarity loss. Specifically, the similarity loss is defined as
\begin{equation}
    {\mathcal{L}}_{sim}(I_s, I_t, S_s, S_t) = 1-NCC(I_t\odot{S_t}, (I_s\odot{S_s})\circ\varphi^{-1})\,,
\end{equation}
where $\odot$ denotes element-wise multiplication.

We use a diffusion regularizer~\cite{modersitzki2003numerical} to encourage smooth displacement fields $u$:
\begin{equation}\label{equ:loss_reg}
    {\mathcal{L}}_{reg}(u) = \frac{1}{|\Omega|}\sum||Du||_F^2\,,
\end{equation}
where $|\Omega|$ is the total volume of the domain and $D$ denotes the Jacobian and $\|\cdot\|_F^2$ denotes the squared Frobenius norm.

\section{Experiments}

\subsection{Data preparation}
We use COPDGene and DirLab for training and evaluation respectively.

{\bf COPDGene} The COPDGene dataset~\cite{regan2011genetic} collects inspiratory and expiratory chest CT scans for 10,000 patients using multi-detector CT scanners on full inspiration (200mAs) and at the end of normal expiration (50mAs). We use a subset of 120 inhale/exhale CT image pairs of the COPDGene dataset with given segmentation masks. We resample the CT images to isotropic spacing of $2.2 mm$ and crop or pad evenly along each dimension to obtain an $160\times{160}\times{160}$ volume with the lung ROI approximately in the center. We compute the associated DRRs via the DRR generator proposed in~\cite{tian2020fluid} with the pre-defined geometry as described in Sec.~\ref{sec:problem_formulation}. The DRRs are computed based on the entire CT image containing the ROI and the surrounding tissues. The resulting DRRs have a size of $200\times{200}$. After processing, the training dataset contains 120 sets of $(I_s, I_t, \textbf{P}_t)$, among which 100 pairs are used for training and 20 pairs are used for validation.  

{\bf DirLab} The DirLab dataset \cite{castillo2013reference} contains 10 pairs of inspiration-expiration lung CT images. Each CT image comes with 300 anatomical landmarks that are manually identified and registered by an expert in thoracic imaging. We applied the same processing to DirLab as for COPDGene and obtain 10 pairs of $(I_s, \textbf{P}_t)$. The segmentation maps of all scans are computed automatically\footnote{https://hub.docker.com/r/acilbwh/chestimagingplatform}. Of note, during evaluation, the 3D volume $I_t$ is not used.

\subsection{Evaluation metrics}
We use the following two evaluation metrics for all our experiments.

{\bf Mean Target Registration Error (mTRE)} This metric computes the mean distance of corresponding landmarks between the warped and the target image. The reported mTRE in our experiments are the mean mTRE of all 10 pairs of DirLab in millimeters. The landmarks are annotated inside the lung based on the lung vessel structure~\cite{castillo2013reference}. This metric is used to evaluate the spatial accuracy of the registration method for fine structures (e.g. lung vessels).

{\bf DICE} The Dice's coefficient measures how much overlap exists between two sets. We compute it between the warped segmentation mask and the target segmentation mask to measure how similar the ROI shapes are after registration.

\subsection{Validation of the DVF subspace}
We compute the gold standard DVFs for the COPDGene dataset via an in-house trained CNN for 3D/3D deformable registration. We construct the subspace by applying PCA to the gold standard DVFs of our COPDGene-train dataset. We form the basis for the deformation space using the eigenvectors which cover $99\%$ of the data variance. This results in $N_e=56$ basis vectors for the DVF subspace. We project the original DVF into this subspace. Then we compare the mTRE, DICE and $\%|J|_{<0}$ between the original DVF and the reconstructed DVF. $\%|J|_{<0}$ denotes the percentage of voxels having negative Jacobian determinant of $\varphi^{-1}$, i.e. exhibiting folding which represents an unrealistic deformation. Tab~\ref{tab:exp_subspace} shows the results. Though the DICE scores decrease significantly, the landmark error only worsens by $2mm$. This is within acceptable range considering the before-registration mTRE is $23.36mm$. The mTRE and DICE score of the reconstructed DVF can be regarded as an upper-bond of the DVF subspace.

\begin{table}[htp]
\vspace{-3mm}
\centering
\begin{tabular}{l|ccc|ccc}
\hline
 & \multicolumn{3}{c|}{Original DVF}& \multicolumn{3}{c}{Reconstructed DVF}      \\ \cline{2-7} 
 & mTRE[mm]$\downarrow$ & DICE[\%]$\uparrow$& $\%|J|_{<0}\downarrow$ & mTRE[mm]$\downarrow$ & DICE[\%]$\uparrow$& $\%|J|_{<0}\downarrow$ \\ \hline
COPDGene-train & - & 98.72 &0.046  & - & 96.46 & 0.020   \\  
COPDGene-val & - & 98.85 & 0.047 & - & 90.90&  0.011  \\
DirLab &3.24/23.36&	98.82 & 0.038  & 5.34/23.36 & 91.88 &0.007 \\\hline
\end{tabular}
\vspace{0.2cm}
\caption{\textbf{Validation of the DVF subspace.} This table shows mTRE, DICE and $\%|J|_{<0}$ on the original DVF and using the reconstructed DVF from the built subspace.}\label{tab:exp_subspace}
\vspace{-12mm}
\end{table}

\subsection{Pairwise 2D/3D deformable image registration}
We conduct an ablation study on the Lift3D module and compare our LiftReg method with other existing approaches for pairwise 2D/3D DIR~\cite{tian2020fluid,zhang2021unsupervised}.

\textbf{Ablation study.} To study how the Lift3D module affects the result, we train a network which concatenates the 2D images $\textbf{P}_t$ and repeats them along the depth direction so that we obtain a 3D volume of size $200\times160\times200$. Then we resample it to the same size as $I_s$ and concatenate it with $I_s$. The rest of the network including $\alpha$ estimation and loss are the same. We denote this network structure as LiftReg-noLift3D. Tab.~\ref{table:exp_existing_methods} shows the results of LiftReg and LiftReg-noLift3D. The landmark error and DICE score of the segmentation masks are improved from $13.89~mm$ to $12.74~mm$ and from 83.69\% to 85.70\% respectively by adding the Lift3D module. This demonstrates that Lift3D provides better spatial information than concatenation.

\textbf{Comparison with existing methods.}
We compare with an optimization-based method~\cite{tian2020fluid} which utilizes large deformation diffeomorphic metric mapping (LDDMM) as the transformation model and optimizes over it using the 2D similarity measure. We use the official public code from this work and run it on the DirLab dataset. We also compare to 2D3D-regnet which uses a UNet as the backbone to predict a DVF given $(I_s, \textbf{P}_t)$. Since the implementation of 2D3D-regnet is not published, we implement it in PyTorch and adapt it to use the s-DCT geometry. To conduct a fair comparison, we train 2D3D-regnet and LiftReg on COPDGene-train and pick the best model based on the validation score of COPDGene-val. We use the NCC similarity loss as the validation score.

Tab.~\ref{table:exp_existing_methods} shows that LiftReg improves the DICE score from 73.55\% to 85.70\% and reduces the mTRE for landmarks inside the lung from $23.36~mm$ to $12.74~mm$. LiftReg outperforms the approach in~\cite{tian2020fluid} and 2D3D-regnet on both metrics. Compared to the learning-based method 2D3D-regnet, LiftReg shows a stronger improvement for mTRE than DICE. One possible explanation could be that the 2D similarity loss contains enough information to guide the learning of the coarse structure such as the boundary of the lung. Hence, while the 3D similarity measure does not results in a significant performance boost for the registration of the lung mask it improve the alignment of fine structures (e.g. lung vessels). 

\begin{table}[htp]
\vspace{-3mm}
\centering
\begin{tabular}{l|ccccc}
\hline
\multicolumn{1}{c|}{Method} & \multicolumn{1}{c}{mTRE[mm]\ $\downarrow$} & \multicolumn{1}{c}{X[mm]\ $\downarrow$} & \multicolumn{1}{c}{Y[mm]\ $\downarrow$} & \multicolumn{1}{c}{Z[mm]\ $\downarrow$} &\multicolumn{1}{c}{DICE[\%]$\uparrow$\ }\\ \hline
Before registration & 23.36$\pm{6.00}$ & 3.97$\pm{1.16}$ &15.37$\pm{7.48}$ & 13.72$\pm{5.22}$ & 73.55$\pm{8.36}$ \\  
Tian et al~\cite{tian2020fluid} &17.39$\pm{4.46}$ &	3.79$\pm{1.09}$ & 12.72$\pm{5.12}$ &	8.57$\pm{2.94}$ &	83.11$\pm{5.61}$   \\
2D3D-regnet &16.94$\pm{4.39}$ & 3.58$\pm{0.96}$ & 9.63$\pm{4.76}$ & 10.71$\pm{3.98}$ & 84.58$\pm{6.04}$   \\ \hline
LiftReg-noLift3D &13.89$\pm{3.63}$ & 3.51$\pm{1.16}$ & 7.978$\pm{3.66}$ & 8.39$\pm{3.67}$ & 83.69$\pm{3.62}$ \\
LiftReg & \textbf{12.74$\pm{3.10}$} &	\textbf{3.43$\pm{1.10}$} & \textbf{7.63$\pm{2.93}$} & \textbf{7.43$\pm{2.15}$} & \textbf{85.70$\pm{3.30}$} \\ \hline
\end{tabular}
\vspace{0.2cm}
\caption{\textbf{Evaluation on DirLab.} Metrics on DirLab for LiftReg and existing methods. X,Y,Z represent the mean absolute landmark error along each axis depicted in Fig.~\ref{fig:DRR_geometry}. From the perspective of CT image orientation, X is the left-right, Y is anterior-posterior, and Z is the superior-inferior direction.}\label{table:exp_existing_methods}
\vspace{-14mm}
\end{table}

\section{Conclusion}
We proposed LiftReg which predicts the basis coefficients of a constructed DVF subspace given a 3D volume and a set of 2D projection images. We demonstrated that LiftReg with statistical deformation model, 3D similarity loss, and Lift3D module outperforms existing pairwise 2D/3D DIR methods in a dataset containing large deformations. A limitation of our method is that it assumes the DRRs are equivalent to real radiographs. This might not be true if we consider scattering effects, beam hardening, and veiling glare~\cite{staub2013digitally}. Interesting future work would be to combine image translation with DRR generation to improve the realism of DRRs with respect to real radiographs; to add support for dynamic geometry where the emitter number $N$ and emitter positions can vary; and to explore more advanced statistical deformation models.

\section{Acknowledgement}
Thanks to Peirong Liu (UNC), Dr. Rong Yuan (Peking University), and Boqi Chen (UNC) for providing valuable suggestions on during the writing of the manuscript. The research reported in this publication was supported by the National Institutes of Health (NIH) under award numbers NIH 1 R01 HL149877 and NIH 1 R01 EB028283. The content is solely the responsibility of the authors and does not necessarily represent the official views of the NIH. 

\bibliographystyle{splncs04}
\bibliography{reference}

\newpage
\appendix
\clearpage
\section{Implementation details}\label{sec:implementaion}

We resample the 3D input images to $160\times160\times160$ to reduce memory consumption. We use $N=4$ emitters evenly distributed over a scanning angle of $30^\circ$. The $\alpha$ estimation module uses seven consecutive convolution blocks with filter dimensions set to 16, 32, 32, 32, 32, 32 respectively. They are followed by three fully connected layers with feature dimensions of 800, 256 and the subspace dimension $N_e=56$. The network is implemented in PyTorch and trained on an NVIDIA RTX A6000 with 48Gb of GPU memory. We set the learning rate to $0.001$, batch size to 30 and $\lambda$ to $0.01$. We train the network for 300 epochs and pick the model which has the best score on the validation dataset during the 300 epochs. We compute $1-NCC(\cdot,\cdot)^2$ between the warped 3D image $I_s\circ\varphi^{-1}$ and $I_t$ as the validation score.

\begin{figure}[htp]
    \centering
    \includegraphics[width=0.8\linewidth]{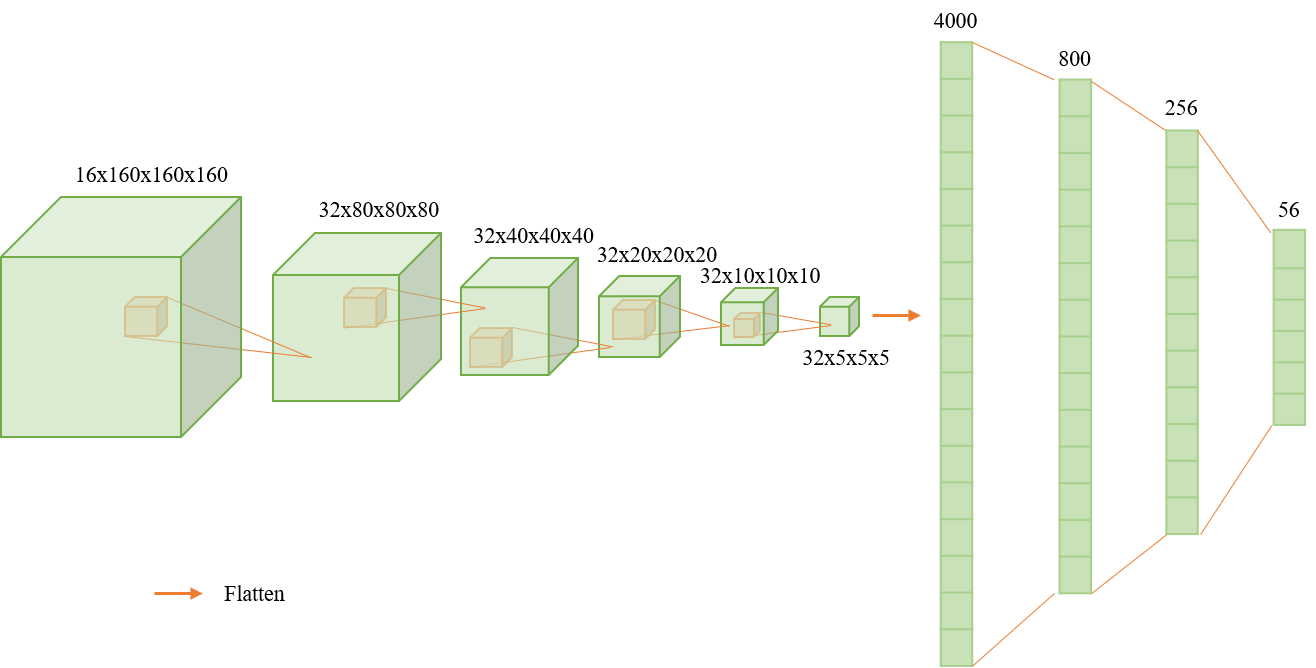}
    \caption{\bf{$\alpha$ estimation module structure.}}
    \label{fig:my_label}
    \vspace{-0.5cm}
\end{figure}

\section{Average execution time}

Tab.~\ref{tab:times} lists the the average runtimes for our different DirLab registration approaches. For the optimization-based method~\cite{tian2020fluid}, the runtime is computed as the elapsed time between the beginning and the end of the optimization process. For the learning-based methods 2D3D-regnet~\cite{zhang2021unsupervised} and LiftReg, the execution time is the inference time.

\begin{table}[b]
\centering
\begin{tabular}{c|cc}
\cline{1-2}
Methods & Time[s] &\\ \cline{1-3}
Tian et al~\cite{tian2020fluid} & 2992.15 &\\
2D3D-regnet & 3.58  &\\
LiftReg & 0.81 &\\ \cline{1-3}
\end{tabular}
\vspace{3mm}
\caption{\bf{Execution time for DirLab.}}
\label{tab:times}
\end{table}

\end{document}